\documentclass[aps,prb,twocolumn,showpacs,amsmath,amssymb,floatfix]{revtex4}
\usepackage{graphicx}% Include figure files
\usepackage{dcolumn}% Align table columns on decimal point
\usepackage{bm}% bold math
\begin{document}
%
%\preprint{JPS/2004March}
%
\title{
Rietveld analysis and maximum entropy method of powder diffraction 
for bundles of single-walled carbon nanotubes
}

\author{Hiroaki Kadowaki}
%\email{kadowaki@comp.metro-u.ac.jp}
\affiliation{Department of Physics, Tokyo Metropolitan University, 
Hachioji-shi, Tokyo 192-0397, Japan}

\author{Akihito Nishiyama}
\affiliation{Department of Physics, Tokyo Metropolitan University, 
Hachioji-shi, Tokyo 192-0397, Japan}
 
\author{Kazuyuki Matsuda}
\affiliation{Department of Physics, Tokyo Metropolitan University, 
Hachioji-shi, Tokyo 192-0397, Japan}
 
\author{Yutaka Maniwa}
\altaffiliation[Also at ]{CREST, JST 
(Japan Science and Technology Corporation).}
\affiliation{Department of Physics, Tokyo Metropolitan University, 
Hachioji-shi, Tokyo 192-0397, Japan}

\author{Shinzo Suzuki}
\affiliation{Department of Chemistry, Tokyo Metropolitan University, 
Hachioji-shi, Tokyo 192-0397, Japan}

\author{Yohji Achiba}
\affiliation{Department of Chemistry, Tokyo Metropolitan University, 
Hachioji-shi, Tokyo 192-0397, Japan}

\author{Hiromichi Kataura}
\affiliation{Nanotechnology Research Institute, 
National Institute of Advanced Industrial Science and Technology, 
Tsukuba, Ibaraki 305-8562, Japan}

\date{\today}

\begin{abstract}
The structure of bundles of single-walled carbon nanotubes (SWNT) 
has been refined by Rietveld analysis using neutron and X-ray 
powder diffraction data. 
Based on previous simulation studies of powder diffraction 
data of SWNT and standard Rietveld analyses, 
we have developed a pattern fit technique 
for SWNT which provides precise structure parameters. 
We also show that the present technique can be 
used with the maximum entropy method (MEM), 
which is complementary to the Rietveld analysis. 
Using the neutron diffraction data of pristine SWNT, 
we have successfully reconstructed the density of carbon 
nuclei and zero density in the inner cavity of SWNT by MEM. 
\end{abstract}

%\pacs{61.46.+w, 61.12.Ld, 68.43.Fg, 81.07.De}

%\keywords{carbon nanotube, neutron diffraction, Rietveld analysis, MEM}

\maketitle

%%%%%%
\section{\label{sec:intro}
Introduction}
In recent years, adsorption of atoms and molecules in bundles of 
single-walled carbon nanotubes~\cite{Iijima93,Bethune93} (SWNT) is attracting 
interests for both fundamental and applied scientists,~\cite{Calbi01} 
ranging from investigations of quasi-one-dimensional phases to 
applications of gas storage.~\cite{Pederson92} 
SWNTs are self-organized into a bundle with a two dimensional (2D) 
triangular lattice structure.~\cite{Thess96} 
In this bundle there are three main adsorption sites: 
an inner cavity of SWNT (tube or t), 
an interstitial channel (IC) between three SWNTs, 
a groove separating two SWNTs on the external surface of 
the bundle [see Fig.~\ref{fig:structure}(a)].~\cite{Calbi01} 
For a small adsorbate, gas uptake may occur progressively 
on the three sites and finally covering the outside of the bundle 
as pressure is increased. 
Experimentally, gas uptakes can be studied by measuring adsorption 
isotherms. 
A more direct method of studying the adsorption sites, 
particularly useful for the IC and tube sites, 
is a diffraction experiment and its structure analysis of 
the 2D triangular lattice. 
Since the diffraction pattern is not affected by uptakes by 
impurities, such as graphite and nano-particles of graphite, 
the diffraction study provides 
complementary information to the adsorption isotherm. 

Techniques of X-ray and neutron diffraction were used to 
demonstrate the triangular lattice of the bundles of 
SWNT,~\cite{Thess96} and to measure the diameter distribution 
of SWNT.~\cite{Rols99,Abe03} 
Powder diffraction patterns by bundles of SWNT with several 
adsorbates have been studied to determine the adsorption 
sites for 
O$_2$ and N$_2$,~\cite{Maniwa99,Fujiwara01} 
CD$_4$,~\cite{Muris02} 
Ar,~\cite{Bienfait03} 
D$_2$,~\cite{Challet03} 
and I$_{n}$.~\cite{Bendiab04} 
As pointed out in Ref.~\onlinecite{Maniwa99}, 
intensity of the (10) reflection increases 
or decreases depending on adsorption at the IC or tube sites, 
respectively, being used as a simple experimental 
distinction. 

When internal structures of adsorbed atoms or molecules 
in the tube site are studied, 
entire powder diffraction patterns have to be analyzed more carefully. 
Among a few examples of this kind, C$_{60}$ and H$_2$O molecules in SWNT 
were shown to form interesting one-dimensional lattices 
in the inner cavity channel using X-ray powder 
diffraction.~\cite{Abe03,Maniwa02,Maniwa05} 
Adsorption of iodine in SWNT and its clustering in the tube and IC sites 
were studied using both X-ray and neutron powder 
diffraction.~\cite{Bendiab04}
However the analyses of these powder diffraction patterns 
are in a primitive stage in comparison with recent 
elaborate techniques of the Rietveld analyses 
for three dimensional (3D) crystals.~\cite{IzumiRIETAN00}
The major difficulty for SWNT is brought about by 
randomness existing in bundles 
which is not controlled even by 
recent preparation methods. 
A bundle consists of different SWNTs with numerous chiralities, 
which are only characterized by a radius distribution with 
an average $R \sim 7$ {\AA} 
and a width $\Delta R \sim 1$ {\AA} 
[full width at half maximum (FWHM)]. 
In addition, radii of bundles in a sample vary 
in a typical range from $\sim 2R$ to $\sim 20R$. 
Small diameter bundles give rise to large broadening of Bragg 
peaks, which complicates the pattern fit method. 

The first purpose of the present work 
is to extend the previous simulation-type studies of 
powder diffraction for SWNT~\cite{Thess96,Rols99,Abe03,Bendiab04} 
to a quantitative Rietveld fit technique. 
We tried to provide a robust basis 
to this Rietveld analysis, which will facilitate to extract 
maximal information of the structure of adsorbates. 
In discussing this analysis in detail, it becomes 
clear that the powder diffraction patterns are more 
informative than previously expected. 
This leads us to another extension of the simulation studies, 
a possibility to exploit a Maximum entropy 
method~\cite{Gull78,Collins82} (MEM) for SWNT.
Thus our second purpose is to formulate the MEM for SWNT, 
and to show its simplest application. 
This MEM can reconstruct the density of carbon and adsorbates 
at the tube site without assuming any structure models, 
which can be a complementary method to the Rietveld analysis, 
where a model structure is assumed from the beginning. 
In this paper, we restrict ourselves to methodologies and 
to analyzing powder diffraction data of pristine SWNT 
samples with no nominal adsorption, showing simple 
demonstrations of the Rietveld analysis and MEM. 

%%%%%
\section{\label{sec:exp}
Experimental}
The raw soot of SWNT was prepared by laser 
vaporization~\cite{Thess96} of a carbon 
rod including Ni and Co catalysts, 
and then purified as previously reported.~\cite{Kataura01} 
The average diameter of the SWNT used was 13.6 {\AA} 
close to 13.7 {\AA} of the (10,10) SWNT. 
To remove adsorbed gases, the sample of SWNT was heated at 
$\sim$ 100$^{\circ}$C 
in an evacuated aluminum container. 
A neutron diffraction experiment was performed using the 
triple-axis spectrometer 4G-GPTAS installed at JRR-3M 
JAERI (Tokai). 
The spectrometer was operated in a triple-axis configuration with 
collimations 30$'$--80$'$--80$'$--open.  
Neutrons of the wave length $\lambda = 2.44$ {\AA} were selected by 
the pyrolytic-graphite (002) monochromator and analyzer. 
Higher-order neutrons were removed by the pyrolytic-graphite filter. 

%%%%%
\section{\label{sec:structureM}
Structure model}
\begin{figure}
\begin{center}
\includegraphics[width=5.0cm,clip]{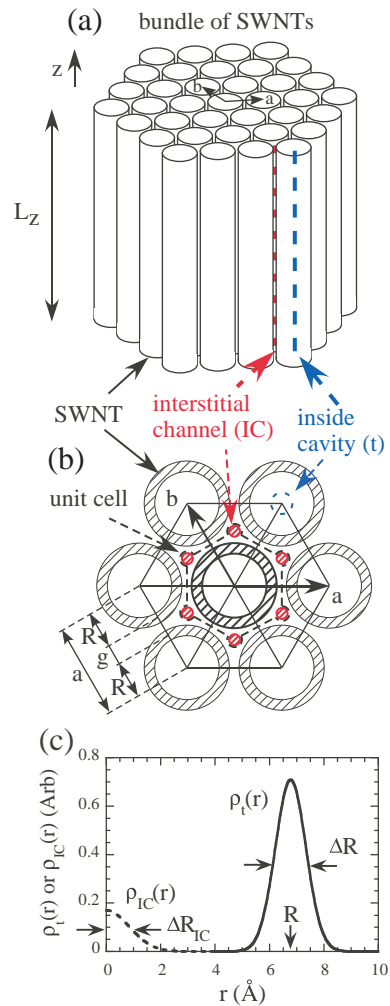}
\end{center}
\caption{\label{fig:structure}
(Color online) 
Structure model of a bundle of SWNTs for analyses of 
powder diffraction experiments. 
SWNTs are approximated by uniform cylinders, in which 
the network structure of the graphene sheet is neglected. 
They crystallize into the 2D triangular lattice as shown 
in (a), where the interstitial channel (IC) and the 
inside cavity of SWNT (t) are shown by dashed lines. 
The unit cell of the triangular lattice is 
the region surrounded by the dashed line in (b), 
where the density of the six ICs are multiplied by a weight 
factor of $\frac{1}{3}$ in calculating the structure factor.
The nuclear densities weighted by scattering lengths 
of nuclei at the tube and IC sites, 
Eqs.~(\ref{eq:NDensityC}) and (\ref{eq:NDensityIC}) with 
parameters of the Rietveld fit result in Sec.~\ref{sec:RietveldA}, 
are illustrated in (c), 
where $r$ represents radius from the tube or IC centers. 
}
\end{figure}

A simple structure model~\cite{Thess96,Rols99,Abe03} 
of a bundle of SWNTs for analyses 
of diffraction experiments is the 2D triangular
lattice of cylinders which is shown in Fig.~\ref{fig:structure}(a). 
In what follows, we use the terminology of neutron diffraction 
for clarity, which can be easily translated into that of 
X-ray diffraction. 
The cylinder represents the average nuclear density of 
carbon nuclei of SWNT in a bundle. 
The average density is regarded to have cylindrical symmetry, 
where atomic network structures of the graphene sheets are ignored, 
because SWNTs in the bundles are assumed to 
have various chiralities 
and to be self-organized without azimuthal correlation
owing to the weak coupling by 
the van der Waals force.~\cite{Thess96} 

We assume that the nuclear density weighted by the scattering length 
of carbon nuclei $b_{\text{C}}$ is well approximated by a Gaussian 
distribution with an average radius $R$ and 
a standard deviation $\sigma_{R}$ 
(FWHM is $\Delta R = \sqrt{8 \ln 2} \sigma_{R}$),
which is illustrated in Fig.~\ref{fig:structure}(c). 
The density around the tube site 
$\rho_{\text{t}}(r)=\rho_{\text{C}}(r)$ is given by 
\begin{equation}
\label{eq:NDensityC}
\rho_{\text{C}}(r) 
=
b_{\text{C}} \sigma_{\text{C}}
\frac{1}{\sqrt{2 \pi} \sigma_{R} }
\exp \left[ - \frac{1}{2} \left( \frac{r-R}{\sigma_{R}} \right)^2 \right],
\end{equation}
where $r$ and $\sigma_{\text{C}}$ 
stand for the radius of the cylindrical coordinates 
and the surface density of carbon nuclei of 
the graphene sheet, respectively.
This uniform cylinder model ($\Delta R \rightarrow 0$ is 
usually assumed) on the 
triangular lattice was proposed to account for images of 
transmission electron microscope and X-ray powder diffraction 
patterns.~\cite{Thess96} 
The fact that the atomic network can be neglected was 
justified by numerical calculations of powder diffraction patterns 
in a low wave-vector range of $Q \lesssim 2$ {\AA}$^{-1}$, 
which were performed for SWNT with single chirality.~\cite{Rols99}

The nano-scale channels in the bundle 
denoted by the tube (t) and IC sites are illustrated in 
Figs.~\ref{fig:structure}(a) and (b). 
For calculations of diffraction patterns, 
adsorbed atoms in the tube site can be treated by an additional 
contribution to $\rho_{\text{t}}(r) = 
\rho_{\text{C}}(r) + \rho_{\text{ads}}(r)$. 
These atoms can show various structures and states 
in the wide inside space of SWNT, which 
should be studied by diffraction experiments.~\cite{Abe03,Maniwa02,Maniwa05} 
On the other hand, adsorbates in the IC site are localized in 
the narrow space surrounded by the three SWNTs. 
For the analysis of powder diffraction in the range 
$Q \lesssim 2$ {\AA}$^{-1}$, 
we can safely assume that the nuclear density is 
well approximated by a Gaussian function centered at the 
IC site with a small width $\sigma_{R_{\text{IC}}}$ 
(half width at half maximum is $\Delta R_{\text{IC}} = 
\sqrt{2 \ln 2} \sigma_{R_{\text{IC}}} = 1$ {\AA}). 
This nuclear density weighted by 
the scattering length of nuclei in the IC site $b_{\text{IC}}$, 
which is shown in Fig.~\ref{fig:structure}(c), 
is written by 
\begin{equation}
\label{eq:NDensityIC}
\rho_{\text{IC}}(r) 
=
b_{\text{IC}} \rho_{\text{IC}}
\exp \left[ - \frac{1}{2} 
\left( \frac{r}{\sigma_{R_{\text{IC}}}} \right)^2 \right],
\end{equation}
where $r$ and $\rho_{\text{IC}}$ are the radius from the IC center 
and the peak nuclear density, respectively. 

The unit cell of the 2D triangular lattice, which is used in 
calculating the structure factor, is depicted by the dashed line 
in Fig.~\ref{fig:structure}(b). 
It should be noted that, to keep the six-fold symmetry of 
the structure factor as a function of 
wave vector, 
we choose a non-standard unit cell 
containing one tube site and six IC sites, 
the latter of which is weighted by $\frac{1}{3}$ for the 
structure factor calculation 
(cf. Eqs.~(\ref{eq:NDensityU}), (\ref{eq:SF2}), 
and (\ref{eq:SF3}) in the appendix).

The average radius $R$ of SWNT in the bundle depends on its 
condition of the preparation, e.g. the local temperature of 
the furnace, 
implying that $R$ takes a number of values for bundles 
in a sample. 
Following 
Refs.~\onlinecite{Abe03} and \onlinecite{Rols99}, 
we assume that the 
probability density of $R$ is approximated by a Gaussian distribution 
with an average $R_{\text{av}}$ and a standard deviation 
$\sigma_{R_{\text{av}}}$ (FWHM is $\Delta R_{\text{av}} = 
\sqrt{8 \ln 2} \sigma_{R_{\text{av}}}$), expressed by 
\begin{equation}
\label{eq:pR}
p(R) =
\frac{1}{\sqrt{2 \pi} \sigma_{R_{\text{av}}} }
\exp \left[ 
- \frac{1}{2} \left( \frac{R-R_{\text{av}}}{\sigma_{R_{\text{av}}}} \right)^2 
\right] .
\end{equation}
We note that the powder diffraction intensity by a bundle, which is 
calculated using a single value of $R$, has to be further 
averaged over the probability density. 
In this average, $R$ dependence of the intertube gap 
$g = a - 2 R$ [see Fig.~\ref{fig:structure}(b)], 
where $a$ is the lattice constant of the triangular lattice, 
can be neglected.~\cite{Abe03}

%%%%%
\section{\label{sec:RietveldA}
Rietveld analysis}
In standard Rietveld analyses~\cite{IzumiRIETAN00} of powder 
diffraction data of 3D crystals, 
the intensity of the diffracted beam at a scattering angle 
$2 \theta$ is fit to 
\begin{equation}
\label{eq:Rietveld}
\begin{split}
I(2 \theta) = 
&s A(2 \theta) 
\sum_{K} m_K |F_K|^2 L(2 \theta_K) 
\Phi (2 \theta - 2 \theta_{K}) \\
&+ y_{\text{b}}(2 \theta) ,
\end{split}
\end{equation}
where $m_K$, $F_K$ and 
$L(2 \theta_K)=1/(\sin \theta_K \sin 2 \theta_K)$ 
are the multiplicity, the structure factor, 
and the Lorentz factor, respectively, 
of the $K$-th reflection with the scattering 
angle $2 \theta_K$; 
$s$, $A(2 \theta)$ and $y_{\text{b}}(2 \theta)$ represent the 
scale factor, the absorption factor, and the background, 
respectively. 
The profile function $\Phi (2 \theta - 2 \theta_{K})$, which is 
normalized by $\int \Phi (2 \theta - 2 \theta_{K}) d (2 \theta) = 1$, 
shows $K$-dependent broadening due to the instrumental resolution and 
the sample crystallinity. 
The functional form of $\Phi$ is often approximated by 
the pseudo-Voigt function, which is further modified to an 
asymmetric form.~\cite{IzumiRIETAN00,Howard82} 

For a bundle of SWNTs, the scattering cross-section 
due to the structure model in the previous section, which 
is derived in the appendix [cf. Eq.~(\ref{eq:CS3})], 
is given by 
\begin{equation}
\label{eq:RietveldCST}
\begin{split}
\frac{d \sigma}{d \Omega} \left( \bm{Q} \right)
\simeq
& L_{z} N_{\text{t}} \frac{16 \pi^3}{\sqrt{3} a^2}
\sum_{\bm{G}} 
\delta(Q_{z}) 
\Hat{\delta} 
\left( \bm{Q}_{\perp} - \bm{G} \right) \\
& \times \left| 
F(\bm{Q}_{\perp})
\right|^{2} , 
\end{split}
\end{equation}
where $L_{z}$ and $N_{\text{t}}$ 
stand for the length of SWNT [see Fig.~\ref{fig:structure}(a)] 
and the number of SWNTs in the bundle, respectively. 
The wave vector 
$\bm{Q}=Q_{z} \bm{e}_z + \bm{Q}_{\perp}$ is 
decomposed into the $z$ component $Q_{z}$ parallel to the direction 
along SWNT, and its perpendicular components 
$\bm{Q}_{\perp} = h \bm{a}^{*} + k \bm{b}^{*}$.
The structure factor $F(\bm{Q}_{\perp})$ is defined by 
Eq.~(\ref{eq:SF}), and the function 
$\Hat{\delta} \left( \bm{Q}_{\perp} - \bm{G} \right)$ 
is the broadened 2D $\delta$ function peaked at a 
reciprocal lattice point $\bm{G}$ of the triangular lattice. 
Simulations of powder diffraction by bundles of SWNTs have been 
performed to study X-ray and neutron diffraction 
data.~\cite{Thess96,Maniwa99,Maniwa01,Maniwa02,Maniwa05,Abe03,Bienfait03,
Rols99,Bendiab04,Challet03,Muris02} 
As noticed by Thess \textit{et al.},~\cite{Thess96} 
an important difference of Eq.~(\ref{eq:RietveldCST}) from 
that of the 3D crystals is the 
broadening of the 2D $\delta$ function by the finite 
size effect of the bundle radius. 
This broadening of the Bragg peaks is much larger than 
instrumental resolutions of 
powder diffractometers. 

Powder averaged intensity of Eq.~(\ref{eq:RietveldCST}), 
corresponding to the first term of Eq.~(\ref{eq:Rietveld}), 
is given by [cf. Eq.~(\ref{eq:CS13})]
\begin{equation}
\label{eq:RietveldT2}
\begin{split}
\Tilde{I}(Q) \simeq 
&s A(Q) 
\sum_{K} m_K |F(Q \Hat{\bm{G}}_{K})|^2 L(Q) \\
& \times \phi (Q - |\bm{G}_{K}|) ,
\end{split}
\end{equation}
where $F(Q \Hat{\bm{G}}_{K})$ is an approximation 
of the powder averaged structure factor 
[cf. Eqs.~(\ref{eq:CS3}), (\ref{eq:CS4}), and (\ref{eq:CS5})], 
and $\phi (Q - |\bm{G}_{K}|)$ is the normalized profile 
function, which is further modified to an 
asymmetric form.~\cite{IzumiRIETAN00,Howard82} 
The use the wave vector $Q=|\bm{Q}|$ instead of scattering 
angle $2 \theta$ is the previous 
convention,~\cite{Thess96,Maniwa99,Rols99} 
where the Lorentz factor becomes 
$L(Q)=1/\sin^2 \theta$ or $\propto 1/Q^2$. 
By further taking the average of Eq.~(\ref{eq:RietveldT2}) 
over the distribution of the radius $R$ of SWNT, Eq.~(\ref{eq:pR}),
we arrive at an expression 
\begin{equation}
\begin{split}
\label{eq:RietveldT1}
I(Q) 
&= \langle \Tilde{I}(Q) \rangle_{R}
+ y_{\text{b}}(Q) \\
\langle \Tilde{I}(Q) \rangle_{R} 
&= \int_{0}^{\infty} p(R) \Tilde{I}(Q) dR , 
\end{split}
\end{equation}
which is used as the calculated intensity 
for the present Rietveld fit. 

\begin{figure}
\begin{center}
\includegraphics[width=8.0cm,clip]{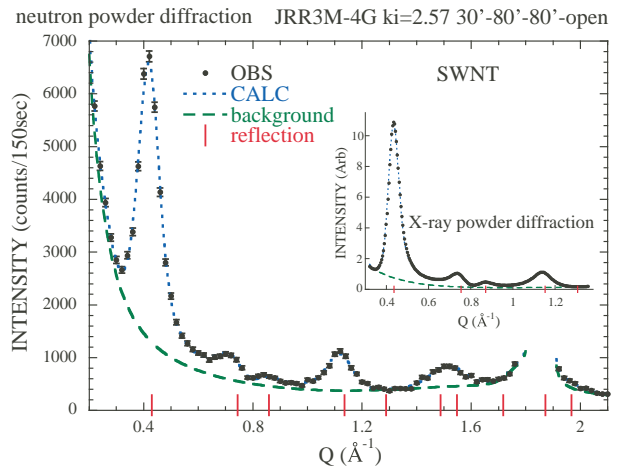}
\end{center}
\caption{\label{fig:PowdPat}
(Color online)
Neutron diffraction pattern of SWNTs at 300 K.
Observed and calculated intensities are plotted by closed circle 
and dotted lines, respectively. 
Background is plotted by dashed line. 
Vertical bars stand for Bragg peak positions with 
$a=2R_{\text{av}}+g$. 
X-ray powder diffraction of the previous work~\cite{Abe03} 
and its Rietveld analysis are shown in the inset. 
}
\end{figure}

Neutron powder diffraction pattern of pristine SWNT 
with no nominal adsorbates measured at 
300 K is shown in Fig.~\ref{fig:PowdPat}. 
The instrumental resolution with the present condition 
is not negligibly small compared to the broadening by the 
finite size effect. 
This additional resolution correction was performed by 
convoluting $\langle \Tilde{I}(Q) \rangle_{R}$ 
of Eq.~(\ref{eq:RietveldT1}) 
with the Gaussian resolution function.~\cite{Cooper67} 
The profile function $\phi (Q - |\bm{G}_{K}|)$ of 
Eq.~(\ref{eq:RietveldT2}) was determined by 
the similar Rietveld fit based on Eq.~(\ref{eq:RietveldT1}) 
using the previous X-ray diffraction 
data~\cite{Abe03} of pristine SWNT. 
These data were measured using a synchrotron X-ray powder 
diffractometer with far higher $Q$ resolution, 
where no resolution convolution was made. 
The profile function could be well approximated by superposition of 
four Gaussian functions. 
The resulting Rietveld analysis is plotted in the inset of 
Fig.~\ref{fig:PowdPat}. 
Using this profile function, the Rietveld analysis of the 
neutron powder diffraction was performed. 
The background $y_{\text{b}}(Q)$ of Eq.~(\ref{eq:RietveldT1}) 
was chosen to be a polynomial in $Q^{-1}$, 
$\sum_{n=0}^{3}c_{n}Q^{-n}$, with 
additional three Gaussian functions centered around 
$Q \sim 1.8$ {\AA}$^{-1}$, where impurity peaks by graphite and 
nano-particles of graphite exist. 
The fitted curve shown in Fig.~\ref{fig:PowdPat} is in 
good agreement with the observed data. 

The fitted parameters of the model structure are: 
$R_{\text{av}} = 6.78 \pm 0.02$ {\AA}, 
$\Delta R_{\text{av}} = 0.56 \pm 0.04$ {\AA} (FWHM), 
$\Delta R = 1.32 \pm 0.08$ {\AA} (FWHM), 
$g = 3.34 \pm 0.04$ {\AA}, 
$\Delta R_{\text{IC}} = 1 $ {\AA} (fixed), 
and 
$b_{\text{IC}} \rho_{\text{IC}}/(b_{\text{C}} \sigma_{\text{C}}) 
= 0.21 \pm 0.08$ {\AA}$^{-1}$, 
for the neutron diffraction data. 
Those for the X-ray diffraction data are: 
$R_{\text{av}} = 6.715 \pm 0.005$ {\AA}, 
$\Delta R_{\text{av}} = 0.68 \pm 0.01$ {\AA} (FWHM), 
$\Delta R = 1.26 \pm 0.03$ {\AA} (FWHM), 
$g = 3.23 \pm 0.01$ {\AA}, 
$\Delta R_{\text{IC}} = 1 $ {\AA} (fixed), 
and 
$\rho_{\text{e,IC}}/ \sigma_{\text{e,C}} 
= 0.11 \pm 0.04$ {\AA}$^{-1}$, 
where $\rho_{\text{e,IC}}$ and $\sigma_{\text{e,C}}$ 
are the peak electron density at the IC site and 
the surface electron density 
of the graphene sheet, respectively. 
These parameters of the X-ray data are consistent with the previous 
work.~\cite{Abe03} 
They also agree reasonably well with the parameters of the 
neutron data, considering the slight sample difference. 
We conclude that the Rietveld fittings have been 
successfully performed for the neutron and X-ray powder 
diffraction data. 

%%%%%
\section{\label{sec:MEM}
Maximum entropy method}
The structural information of the unit cell is contained in 
the structure factor $F(\bm{Q})$. 
In a Rietveld analysis of an ordinary 3D crystal, 
structure factors are observed at a number of 
discrete wave vectors $\bm{G}$. 
On the other hand, there are only 10 reflections in the 
present neutron powder diffraction data of SWNT 
(see Fig.~\ref{fig:PowdPat}). 
From this small number, 
one may think that it is very difficult to determine 
many structural parameters. 
However, this is not simply the case, owing to 
the other aspect of the important difference, i.e., 
the large width of the profile function, 
showing substantial overlaps with 
neighboring reflections, which is easily seen 
in the range $0.4<Q<0.8$ {\AA}$^{-1}$ of 
Fig.~\ref{fig:PowdPat}. 
This overlap suggests a possibility to make use of the 
structure factor in a continuous $Q$ range. 
If the observer range of $0.2<Q<2.1$ {\AA}$^{-1}$ 
is fully available, 
one may think that the powder diffraction data 
contain more information than 
the simple expectation. 
As noticed in Ref.~\onlinecite{Maniwa01}, this is the 
reason why the parameter $R_{\text{av}}$ is precisely determined 
by powder diffraction of SWNT. 
In the following, we discuss a method which takes advantage 
of the availability of the continuous $Q$ range. 

Adsorbates in the IC site, which are localized in the small space, 
cannot have internal structures except for 
those in the direction along SWNT. 
Thus we assume that the IC part of the structure factor 
has only one parameter $\rho_{\text{IC}}$, which 
can be easily determined. 
Then remaining problem is the structure at the tube site, 
SWNT and its inside, which is expressed by $\rho_{\text{t}}(r)$. 
The density $\rho_{\text{t}}(r)$ is related to 
the tube part of the structure factor by [cf. Eq.~(\ref{eq:SF4})] 
\begin{equation}
\label{eq:FB}
F_{\text{t}}(Q)
=
2 \pi \int_{0}^{\infty}
r J_{0}(Q r) \rho_{\text{t}}(r)dr ,
\end{equation}
where $J_{0}(Q r)$ is the cylindrical Bessel function. 
Thus the problem is reduced to how to determine 
$\rho_{\text{t}}(r)$ from $F_{\text{t}}(Q)$. 
Since Eq.~(\ref{eq:FB}) is the Fourier-Bessel transform, 
a solution is to use the inverse transform 
\begin{equation}
\label{eq:FBinv}
\rho_{\text{t}}(r)
=
\frac{1}{2 \pi} \int_{0}^{\infty}
Q J_{0}(Q r)F_{\text{t}}(Q) dQ .
\end{equation}
However an application of this inversion to the neutron diffraction 
data failed, because the inverted $\rho_{\text{t}}(r)$ 
showed a spurious oscillation, which is a cut-off effect 
due to insufficient $Q$ range. 
A well known method to avoid this cut-off effect of 
Fourier transform is MEM,~\cite{Gull78,Skilling84} 
which is being successfully used in single-crystal and 
powder diffraction analyses.~\cite{Collins82,IzumiRIETAN00,Takata95} 

\begin{figure}
\begin{center}
\includegraphics[width=8.0cm,clip]{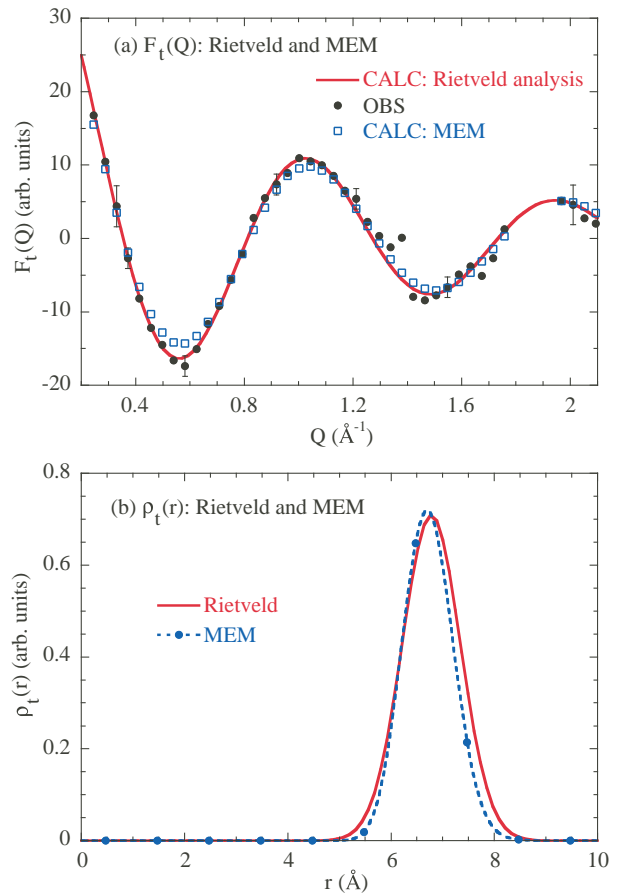}
\end{center}
\caption{\label{fig:MEM}
(Color online)
(a) Structure factors of the tube site $F_{\text{t}}(Q)$ with 
$R = R_{\text{av}}$ are plotted as a function of wave vector. 
The full line is the result of the Rietveld analysis in 
Sec.~\ref{sec:RietveldA}. 
Closed circle represents the observed 
$\left[ F_{\text{t}}(Q_{n}) \right]_{R=R_{\text{av}}}$ 
determined by the fit described in the text. 
Open square is the calculated 
$\left[ F_{\text{t}}(Q_{n}) \right]_{R=R_{\text{av}}}$ 
using Eq.~(\ref{eq:FBdisc}) and $\rho_{\text{t}}(r_{m})$ 
obtained by MEM. 
(b) Nuclear densities at the tube site obtained by 
the Rietveld analysis in Sec.~\ref{sec:RietveldA} with 
$R=R_{\text{av}}$ and MEM. 
}
\end{figure}

In general, MEM is based on a linear relation between 
observed and calculated quantities 
(data and image in the MEM terminology~\cite{Skilling84}). 
This relation is Eq.~(\ref{eq:FB}) for 
$F_{\text{t}}(Q)$ (data) and $\rho_{\text{t}}(r)$ (image). 
To apply a general purpose MEM algorithm,~\cite{Skilling84} 
a discrete version of the linear relation 
\begin{equation}
\label{eq:FBdisc}
F_{\text{t}}(Q_{n})
=
2 \pi \sum_{m}
J_{0}(Q_{n} r_{m}) r_{m} \Delta r_{m} \rho_{\text{t}}(r_{m}) 
\end{equation}
was used as a basis for the present MEM. 
The first step of the MEM is to determine 
observed values of $F_{\text{t}}(Q_{n})$. 
We calculated these by a Rietveld fit, 
where $F_{\text{t}}(Q_{n})$ are fitting parameters 
and intermediate $F_{\text{t}}(Q)$ is evaluated 
by a linear interpolation. 
In this fitting the other parameters: 
the background parameters of $y_{\text{b}}(Q)$, $s$, 
$b_{\text{IC}} \rho_{\text{IC}}$, $\Delta R_{\text{IC}}$, 
$g$, $R_{\text{av}}$, and $\Delta R_{\text{av}}$, 
were fixed to the values of the previous section. 
It should be noted that the three parameters, 
$g$, $R_{\text{av}}$, and $\Delta R_{\text{av}}$, 
and the SWNT radius $R$ still have meaning 
in connection with the lattice constant $a=2R+g$. 
Thus the fitting parameters $F_{\text{t}}(Q_{n})$ 
depend on $R$, and we refer to this $R$ dependence 
explicitly by 
$\left[ F_{\text{t}}(Q_{n}) \right]_{R}$. 
In performing the average over 
$R$ [cf. Eq.~(\ref{eq:RietveldT1})], 
we assumed $R$ dependence 
\begin{equation}
\label{eq:Rdep}
\left[ F_{\text{t}}(Q_{n}) \right]_{R}
= \frac{R}{R_{\text{av}}} 
F_{\text{t}}^{(0)} \left( \frac{R}{R_{\text{av}}}Q_{n} \right) ,
\end{equation}
which is a good approximation for empty SWNT. 
The Rietveld fit was carried out and resulting parameters 
$\left[ F_{\text{t}}(Q_{n}) \right]_{R=R_{\text{av}}}$ 
are shown in 
Fig.~\ref{fig:MEM}(a) together with that 
of the Rietveld analysis in the previous section. 

The second step of the present MEM is to reconstruct 
$\rho_{\text{t}}(r_{m})$ from 
$\left[ F_{\text{t}}(Q_{n}) \right]_{R=R_{\text{av}}}$. 
We carried out this using the general purpose MEM 
algorithm.~\cite{Skilling84} 
The reconstructed $\rho_{\text{t}}(r_{m})$ is plotted 
in Fig.~\ref{fig:MEM}(b), where the Gaussian function 
Eq.~(\ref{eq:NDensityC}) with $R=R_{\text{av}}$ 
determined by the Rietveld analysis in the previous section 
is also shown. 
One can see that the Rietveld and MEM results 
agree surprisingly well. 
The structure factors 
$\left[ F_{\text{t}}(Q_{n}) \right]_{R=R_{\text{av}}}$ 
were calculated 
using the MEM reconstructed $\rho_{\text{t}}(r_{m})$ 
and Eq.~(\ref{eq:FBdisc}), and are plotted in 
Fig.~\ref{fig:MEM}(a). 
They agree well with the observed 
$\left[ F_{\text{t}}(Q_{n}) \right]_{R=R_{\text{av}}}$ 
apart from small discrepancy around 
$Q \sim 0.55$ {\AA}$^{-1}$, 
implying that the present MEM gives self-consistent 
results. 
Therefore we conclude that the continuous $Q$ data 
of structure factor have been excellently analyzed by 
the present MEM formulation. 

%%%%%
\section{\label{sec:dis}
Discussion}
The advantage of MEM~\cite{Gull78,Skilling84} is 
providing a way to choose a maximally 
non-committal nuclear density without any structure 
models among a series of nuclear densities consistent with 
experimental data. 
Thus MEM is a complementary method to the Rietveld analysis, 
in which an empirical structure model is assumed from the beginning. 
Since the formulation of MEM discussed in Sec.~\ref{sec:MEM} 
partly relies on the result of the Rietveld analysis 
of Sec.~\ref{sec:RietveldA}, 
we have to check consistency between the MEM reconstructed 
nuclear density and that of the empirical structure model. 
Fig.~\ref{fig:MEM}(b) definitely warrants this consistency 
for the pristine SWNT sample. 
For general cases, where atoms are adsorbed 
inside SWNT, the consistency is not necessarily obtained after the 
first trial of MEM. 
If the first consistency check is negative, we have to adjust 
the structure model considering the first MEM reconstructed 
nuclear density 
and perform second Rietveld and MEM analyses. 
Depending on the second consistency check, 
we may arrive at a final answer or continue this 
iterative procedure. 
This kind of iterative use of the Rietveld and MEM 
analyses have been skillfully used for powder diffraction 
of 3D crystals.~\cite{IzumiRIETAN00,Takata95} 
Applications of the present Rietveld and MEM analyses to 
SWNT with adsorbates in the tube site are awaited. 

%%%%%
\section{\label{sec:conc}
Conclusions}
We have developed a pattern fit technique, Rietveld analysis, 
for powder diffraction data of 
bundles of SWNTs, 
by extending previous simulation studies. 
Neutron and X-ray powder diffraction data of pristine 
SWNT samples have been successfully analyzed 
by this Rietveld analysis. 
The high fitting quality indicates that the analysis will be 
useful to obtain structure parameters of 
adsorbates in SWNT samples. 
We also have exploited a MEM for reconstructing 
nuclear or electron density at the tube site 
with the use of the pattern fit technique. 
We have applied this MEM to the neutron diffraction data 
and reconstructed the nuclear density of SWNT 
consistent with the Rietveld result. 

%%%%%
%\section*{Acknowledgment}
\begin{acknowledgments}
We are grateful to K. Shibata and H. Yoshizawa for 
valuable discussions.
H. Kataura acknowledges a support by 
Industrial Technology Research Grant Program in '03 
from New Energy and 
Industrial Technology Development Organization (NEDO) of Japan.
\end{acknowledgments}

%%%%%
\appendix*
\section{Scattering cross-section}
The neutron scattering cross-section by a bundle of SWNTs is calculated 
using the general formula~\cite{Lovesey84} 
\begin{equation}
\label{eq:CS1}
\frac{d \sigma}{d \Omega} \left( \bm{Q} \right)
=
\left| 
\int \exp(i \bm{Q} \cdot \bm{r}) 
\rho(\bm{r}) d \bm{r}
\right|^{2} ,
\end{equation}
where $\rho(\bm{r})$ stands for the nuclear density 
weighted by the scattering length of each nucleus. 
On the basis of this equation one can discuss the Bragg scattering, 
scattering from a single SWNT, cluster of atoms or molecules, small 
angle scattering, etc.

For the Bragg scattering due to the model structure 
shown in Fig.~\ref{fig:structure}, 
the nuclear density is expressed by 
\begin{equation}
\label{eq:NDensity}
\begin{split}
\rho(\bm{r}=\bm{r}_{\perp} + z \bm{e}_{z}) 
= &\sum_{\bm{r}_{\perp,\text{t}}} 
\rho_{\text{t}}(|\bm{r}_{\perp} - \bm{r}_{\perp,\text{t}}|) \\
&+ \sum_{\bm{r}_{\perp,\text{IC}}} 
\rho_{\text{IC}}(|\bm{r}_{\perp} - \bm{r}_{\perp,\text{IC}}|). 
\end{split}
\end{equation}
The first term is the nuclear density of SWNT and adsorbates 
at the tube site $\bm{r}_{\perp,\text{t}} = 
n_{a} \bm{a} + n_{b} \bm{b} $. 
The second term is the nuclear density of adsorbates at the IC site 
$\bm{r}_{\perp,\text{IC}} = 
\bm{r}_{\perp,\text{t}}
 \pm \left( \frac{1}{3} \bm{a} + \frac{2}{3} \bm{b} 
\right) $ 
$ \left[ \text{or}
 \pm \left( \frac{2}{3} \bm{a} + \frac{1}{3} \bm{b} \right) 
, \pm \left( \frac{1}{3} \bm{a} - \frac{1}{3} \bm{b} \right) 
\right] $. 
By inserting Eq.~(\ref{eq:NDensity}) in Eq.~(\ref{eq:CS1}) 
and performing the integration along the $z$ axis 
in $0<z<L_{z}$ with large $L_{z} \geq 1$ $\mu$m,~\cite{Thess96} 
one has 
\begin{equation}
\label{eq:CS2}
\begin{split}
\frac{d \sigma}{d \Omega} \left( \bm{Q} \right)
=
&2 \pi L_{z} \delta(Q_{z}) 
\left| 
\sum_{\bm{r}_{\perp,\text{t}}} 
\exp(i \bm{Q}_{\perp} \cdot \bm{r}_{\perp,\text{t}}) 
\right|^{2} \\
&\times \left| F(\bm{Q}_{\perp})
\right|^{2} ,
\end{split}
\end{equation}
where the summation is taken over 2D triangular lattice 
positions. 
The structure factor $F(\bm{Q}_{\perp})$ is given by 
\begin{equation}
\label{eq:SF}
F(\bm{Q}_{\perp})
=
\int_{\text{unit cell}}
\exp(i \bm{Q}_{\perp} \cdot \bm{r}_{\perp}) 
\rho(\bm{r}_{\perp}) 
d \bm{r}_{\perp} ,
\end{equation}
where the unit cell is shown in Fig.~\ref{fig:structure}(b).
We note that although the $\delta$ function in Eq.~(\ref{eq:CS2}) 
is sharp along the $Q_{z}$ direction, 
the lattice sum peaks at reciprocal lattice points 
$\bm{G}$ with a finite width in the $\bm{Q}_{\perp}$ 
space.~\cite{Thess96} 
The latter would become 2D $\delta$ function only in the limit of 
the infinitely large radius of the bundle, as 
\begin{equation}
\label{eq:DF}
\left| 
\sum_{\bm{r}_{\perp,\text{t}}} 
\exp(i \bm{Q}_{\perp} \cdot \bm{r}_{\perp,\text{t}}) 
\right|^{2}
\rightarrow 
N_{\text{t}} 
\frac{8 \pi^2}{\sqrt{3} a^2}
\sum_{\bm{G}} 
\delta(\bm{Q}_{\perp} - \bm{G}) ,
\end{equation}
where $N_{\text{t}} (\rightarrow \infty)$ is the number of SWNTs. 

The finite width of the Bragg-like scattering in 
Eq.~(\ref{eq:CS2}) by small $N_{\text{t}}$ requires us to 
carefully treat the structure factor of Eq.~(\ref{eq:SF}), 
because $F(\bm{Q}_{\perp})$ should be evaluated for continuously 
varying $\bm{Q}_{\perp}$. 
In view of the rope shape of TEM images of bundles,~\cite{Thess96} 
which has quasi-six-fold symmetry, 
it is probably adequate to keep the six-fold symmetry of 
$F(\bm{Q}_{\perp})$ for continuously varying $\bm{Q}_{\perp}$. 
It can be retained for the unit cell shown in 
Fig.~\ref{fig:structure}(b), explicitly using the nuclear density 
\begin{equation}
\label{eq:NDensityU}
\rho(\bm{r}_{\perp})
=
\rho_{\text{t}}(|\bm{r}_{\perp}|)
+
\frac{1}{3} 
\sum_{i=1}^{6} 
\rho_{\text{IC}}(|\bm{r}_{\perp} - \bm{r}_{\perp, \text{IC},i}|)
\end{equation}
in Eq.~(\ref{eq:SF}), where $\bm{r}_{\perp, \text{IC},i}$ is 
the surrounding six IC sites around the tube 
site $\bm{r}_{\perp,t}=0$. 
Substituting Eq.~(\ref{eq:NDensityU}) for Eq.~(\ref{eq:SF}), 
the structure factor becomes 
\begin{equation}
\label{eq:SF2}
F(\bm{Q}_{\perp})
=
F_{\text{t}}(|\bm{Q}_{\perp}|)
+
g(\bm{Q}_{\perp}) F_{\text{IC}}(|\bm{Q}_{\perp}|), 
\end{equation}
where 
\begin{equation}
\begin{split}
\label{eq:SF3}
g(\bm{Q}_{\perp}
&=h \bm{a}^{*} + k \bm{b}^{*}) =
\frac{2}{3} \bigg\{ 
\cos \left[ \frac{2 \pi}{3} (2 h + k) \right] \\
&+ \cos \left[ \frac{2 \pi}{3} (h + 2 k) \right] 
+ \cos \left[ \frac{2 \pi}{3} (h - k) \right]
\bigg\} . 
\end{split}
\end{equation}
The two terms in Eq.~(\ref{eq:SF2}) are the contribution from 
the tube site 
\begin{equation}
\begin{split}
\label{eq:SF4}
F_{\text{t}}(|\bm{Q}_{\perp}|)
&=
\int
\exp(i \bm{Q}_{\perp} \cdot \bm{r}_{\perp}) 
\rho_{\text{t}}(|\bm{r}_{\perp}|) 
d \bm{r}_{\perp} \\
&=
2 \pi \int_{0}^{\infty}
r J_{0}(Q_{\perp} r) \rho_{\text{t}}(r)dr ,
\end{split}
\end{equation}
and that from the IC site 
\begin{equation}
\begin{split}
\label{eq:SF5}
F_{\text{IC}}(|\bm{Q}_{\perp}|)
&=
\int
\exp(i \bm{Q}_{\perp} \cdot \bm{r}_{\perp}) 
\rho_{\text{IC}}(|\bm{r}_{\perp}|) 
d \bm{r}_{\perp} \\
&=
2 \pi \int_{0}^{\infty}
r J_{0}(Q_{\perp} r) \rho_{\text{IC}}(r)dr .
\end{split}
\end{equation}

Considering Eqs.~(\ref{eq:CS2}) and (\ref{eq:DF}), 
the neutron scattering 
cross-section due to the Bragg-like scattering of the 2D 
triangular lattice is summarized as 
\begin{equation}
\label{eq:CS3}
\begin{split}
\frac{d \sigma}{d \Omega} \left( \bm{Q} \right)
\simeq
& L_{z} N_{\text{t}} \frac{16 \pi^3}{\sqrt{3} a^2}
\sum_{\bm{G}} 
\delta(Q_{z}) 
\Hat{\delta} 
\left( \bm{Q}_{\perp} - \bm{G} \right) \\
& \times \left| 
F(\bm{Q}_{\perp})
\right|^{2} , 
\end{split}
\end{equation}
where the function $\Hat{\delta} \left( \bm{Q}_{\perp} - 
\bm{G} \right)$ is a broadened 2D $\delta$ 
function normalized by $\int \Hat{\delta} \left( 
\bm{Q}_{\perp} - \bm{G} \right) 
d \bm{Q}_{\perp} = 1$. 

To calculate intensity of powder diffraction, 
this cross-section is further averaged 
over all orientation of the bundle, or equivalently 
over all direction of $\bm{Q}$,~\cite{Bendiab04} i.e.,
\begin{equation}
\label{eq:CS4}
\begin{split}
\Tilde{I}(Q)
&=
\left\langle \frac{d \sigma}{d \Omega} \left( \bm{Q} \right) 
\right\rangle_{\left| \bm{Q} \right| = Q, \text{powder av.}} \\
&=
\frac{1}{4 \pi Q^2}
\int \frac{d \sigma}{d \Omega} \left( \bm{Q} \right) d\bm{S}_{\bm{Q}} , 
\end{split}
\end{equation}
where $d\bm{S}_{\bm{Q}}$ is the infinitesimal area 
on the sphere of radius $Q$. 
Since the numerical evaluation of this powder average 
took too long time to perform the Rietveld fitting, 
we made an ad hoc approximation given by 
\begin{equation}
\label{eq:CS5}
\begin{split}
\Tilde{I}(Q) \simeq 
& L_{z} N_{\text{t}} \frac{16 \pi^3}{\sqrt{3} a^2} 
\sum_{\bm{G}} 
\phi(Q - | \bm{G} |) \frac{1}{4 \pi Q^2} \\
& \times \left| 
F(Q \Hat{\bm{G}})
\right|^{2} ,
\end{split}
\end{equation}
where a profile function $\phi(Q - G)$ normalized by 
$\int \phi(Q-G) dQ = 1$ is used, and 
$\Hat{\bm{G}}=\bm{G}/G$. 
We checked numerically that Eq.~(\ref{eq:CS5}) is a good 
approximation for the fits of 
the present experimental data. 
By using the six-fold symmetry of the structure factor, 
we finally write Eq.~(\ref{eq:CS5}) in a form 
appropriate for the powder diffraction 
\begin{equation}
\label{eq:CS13}
\begin{split}
\Tilde{I}(Q) \simeq 
& L_{z} N_{\text{t}} \frac{4 \pi^2}{\sqrt{3} a^2} 
\sum_{K} m_{K}
\phi(Q - | \bm{G}_{K} |) \frac{1}{Q^2} \\
& \times \left| 
F(Q \Hat{\bm{G}}_{K}) 
\right|^{2} ,
\end{split}
\end{equation}
where $m_{K}$ is the multiplicity.

\end{document}